\documentclass[fleqn,twoside]{article}
\usepackage[headings]{espcrc2}

\readRCS
$Id: espcrc2.tex,v 1.2 2004/02/24 11:22:11 spepping Exp $
\usepackage{graphicx}
\usepackage[figuresright]{rotating}

\newcommand{\AmS}{{\protect\the\textfont2
  A\kern-.1667em\lower.5ex\hbox{M}\kern-.125emS}}

\def\beq{\begin{equation}}
\def\eeq{\end{equation}}
\def\beqar{\begin{eqnarray}}
\def\eeqar{\end{eqnarray}}
\def\barr#1{\begin{array}{#1}}
\def\earr{\end{array}}
\def\bfi{\begin{figure}}
\def\efi{\end{figure}}
\def\btab{\begin{table}}
\def\etab{\end{table}}
\def\bce{\begin{center}}
\def\ece{\end{center}}
\def\nn{\nonumber}

\def\text{\textstyle}

\def\ga{\gamma}
\def\Ga{\Gamma}

\def\mathswitchr#1{\relax\ifmmode{\mathrm{#1}}\else$\mathrm{#1}$\fi}

\newcommand{\PW}{\mathswitchr W}
\newcommand{\PZ}{\mathswitchr Z}

\newcommand{\PH}{\mathswitchr H}

\newcommand{\Pe}{\mathswitchr e}

\newcommand{\Pne}{\mathswitch \nu_{\mathrm{e}}}

\newcommand{\Pep}{\mathswitchr {e^+}}
\newcommand{\Pem}{\mathswitchr {e^-}}
\newcommand{\Pmum}{\mathswitchr {\mu^-}}

\def\mathswitch#1{\relax\ifmmode#1\else$#1$\fi}

\newcommand{\MZ}{\mathswitch {M_\PZ}}
\newcommand{\MH}{\mathswitch {M_\PH}}

\newcommand{\GeV}{\unskip\,\mathrm{GeV}}
\newcommand{\MeV}{\unskip\,\mathrm{MeV}}

\def\draftdate{\relax}
\def\mda{\relax}
\def\mua{\relax}
\def\mla{\relax}
\def\draft{
\def\thtystars{******************************}
\def\sixtystars{\thtystars\thtystars}
\typeout{}
\typeout{\sixtystars**}
\typeout{* Draft mode!
         For final version remove \protect\draft\space in source file
*}
\typeout{\sixtystars**}
\typeout{}
\def\draftdate{\today}
\def\mua{\marginpar[\boldmath\hfil$\uparrow$]%
                   {\boldmath$\uparrow$\hfil}%
                    \typeout{marginpar: $\uparrow$}\ignorespaces}
\def\mda{\marginpar[\boldmath\hfil$\downarrow$]%
                   {\boldmath$\downarrow$\hfil}%
                    \typeout{marginpar: $\downarrow$}\ignorespaces}
\def\mla{\marginpar[\boldmath\hfil$\rightarrow$]%
                   {\boldmath$\leftarrow $\hfil}%
                    \typeout{marginpar: $\leftrightarrow$}\ignorespaces}
\def\Mua{\marginpar[\boldmath\hfil$\Uparrow$]%
                   {\boldmath$\Uparrow$\hfil}%
                    \typeout{marginpar: $\uparrow$}\ignorespaces}
\def\Mda{\marginpar[\boldmath\hfil$\Downarrow$]%
                   {\boldmath$\Downarrow$\hfil}%
                    \typeout{marginpar: $\downarrow$}\ignorespaces}
\def\Mla{\marginpar[\boldmath\hfil$\Rightarrow$]%
                   {\boldmath$\Leftarrow $\hfil}%
                    \typeout{marginpar: $\leftrightarrow$}\ignorespaces}
\overfullrule 5pt
\oddsidemargin -1mm
\marginparwidth 18mm
}

\title{Precision calculations for the Higgs decays
  $\PH\to\PZ\PZ/\PW\PW\to4$ leptons}

\author{A.~Bredenstein\address[KEK]{High Energy Accelerator Research
                Organization (KEK), Tsukuba, Ibaraki 305-0801, Japan}, 
        A.~Denner\address{Paul Scherrer Institut, W\"urenlingen und Villigen,
        CH-5232 Villigen PSI, Switzerland},
        S.~Dittmaier\address[MPI]{Max-Planck-Institut f\"ur Physik
        (Werner-Heisenberg-Institut),
        D-80805 M\"unchen, Germany}
        and
        M.M.~Weber\address{Bergische Universit\"at Wuppertal, 
        D-42097 Wuppertal, Germany} }
       
\runtitle{Electroweak corrections to the Higgs decays
  $\PH\to\PZ\PZ/\PW\PW\to4$ leptons}
\runauthor{A.~Bredenstein, A.~Denner, S.~Dittmaier and M.M.~Weber}

\begin{document}

\begin{abstract}
  Extending earlier work, we provide predictions obtained with the Monte Carlo
  generator {\sl PROPHECY4f\/} for the decays
  $\PH\to\PZ\PZ/\PW\PW\to4l$ including the complete
  electroweak ${\cal O}(\alpha)$ corrections and some higher-order
  improvements. 
  The gauge-boson resonances are described in the complex-mass scheme. 
  Here, particular attention is paid to a comparison of
  different final states with identical 
charges, such as
  $\Pep\Pem\mu^+\mu^-$ and $\mu^+\mu^-\mu^+\mu^-$.
\vspace{1pc}
\end{abstract}

\maketitle

\vspace*{-4em}

\section{INTRODUCTION}

The primary task of the LHC will be the detection and the study of the
Higgs boson. If it is heavier than $140\,$GeV, it decays dominantly
into gauge-boson pairs, i.e.\ into 4 fermions. These decays offer the
largest discovery potential for a Higgs boson with a mass
$\MH\mathrel{\raisebox{-.3em}{$\stackrel{\displaystyle
      >}{\sim}$}}130\,$GeV \cite{Asai:2004ws}, and the decay
$\PH\to\PZ\PZ\to4l$ will allow for the most accurate measurement of
$\MH$ above $130\,$GeV \cite{Zivkovic:2004sv}.  At an
$\rm{e}^+\rm{e}^-$ linear collider, these decays will enable
measurements of the corresponding
branching ratios and couplings at the few-per-cent level.

A kinematical reconstruction of the Higgs boson and of the virtual W
and Z bosons requires the study of distributions.  Thereby, it is
important to include radiative corrections, in particular real photon
radiation. In addition, the verification of the spin and the CP
properties of the Higgs boson relies on the study of angular and
invariant-mass distributions \cite{Barger:1993wt,Choi:2002jk}. As a
consequence a Monte Carlo generator for
$\PH\to\PZ\PZ/\PW\PW\to4f$ including electroweak corrections is
needed.

In the past, only the electroweak ${\cal O}(\alpha)$ corrections to
decays into on-shell gauge bosons $\PH\to\PZ\PZ/\PW\PW$
\cite{Fleischer:1980ub} and some leading higher-order corrections were
known. However, in the threshold region the on-shell approximation
becomes unreliable.  Below the gauge-boson-pair thresholds only the
leading order was known until recently.

{\sl PROPHECY4f} \cite{Bredenstein:2006rh} is a recently constructed
Monte Carlo event generator for $\PH\to\PZ\PZ/\PW\PW\to4f$ that
includes electroweak corrections as well as some higher-order
improvements. Since the process with off-shell gauge bosons is
consistently considered without any on-shell approximations, the
obtained results are valid above, near, and below the gauge-boson pair
thresholds.  Parallel to our work, another generator for
$\PH\to\PZ\PZ\to4f$, including electromagnetic corrections only,
has been introduced in Ref.~\cite{CarloniCalame:2006vr}.

In this note we briefly describe the structure of {\sl PROPHECY4f\/}
and of the underlying calculations.  Moreover, we extend the numerical
results of Ref.~\cite{Bredenstein:2006rh} obtained with this code
by paying particular attention to final states that
differ only in the generation quantum numbers of the leptons. 

\section{CALCULATIONAL DETAILS}

We have calculated the complete electroweak ${\cal O}(\alpha)$ 
corrections to the processes $\PH\to4f$. This
includes both the corrections to the decays
$\PH\to\PZ\PZ\to4f$ and
$\PH\to\PW\PW\to4f$ and their interference.
The calculation of the one-loop diagrams has been performed in the
conventional 't~Hooft--Feynman gauge and in the background-field
formalism using the conventions of Refs.~\cite{Denner:1991kt} and
\cite{Denner:1994xt}, respectively.  The masses of the external
fermions have been neglected whenever possible, i.e.\ everywhere but
in the mass-singular logarithms.

For the implementation of the finite widths of the gauge bosons we use
the ``complex-mass scheme'', which was introduced in
Ref.~\cite{Denner:1999gp} for lowest-order calculations and
generalized to the one-loop level in Ref.~\cite{Denner:2005fg}.  In
this approach the W- and Z-boson masses are consistently considered as
complex quantities, defined as the locations of the propagator poles
in the complex plane.  The scheme fully respects all relations that
follow from gauge invariance.  A brief description of this scheme can
also be found in Ref.~\cite{Denner:2006ic}.

The amplitudes have been generated with {\sl FeynArts}, using the two
independent versions 1 and 3, as described in
Refs.~\cite{Kublbeck:1990xc} and \cite{Hahn:2000kx}, respectively.
The algebraic evaluation has been performed in two completely
independent ways. One calculation is based on an in-house program
implemented in {\sl Mathematica}, the other has been completed with
the help of {\sl FormCalc} \cite{Hahn:1998yk}.  
The amplitudes are
expressed in terms of standard matrix elements and coefficients of
tensor integrals as described in the appendix of
Ref.~\cite{Denner:2003iy}.

The tensor coefficients are evaluated as in the calculation of the
corrections to ${\rm e}^+{\rm e}^-\to4\,$fermions
\cite{Denner:2005fg,Denner:2005es}.  They are recursively reduced to
master integrals at the numerical level.  The scalar master integrals
are evaluated for complex masses using the methods and results of
Refs.~\cite{'tHooft:1979xw,Beenakker:1990jr,Denner:1991qq}.  UV
divergences are regulated dimensionally and IR divergences with an
infinitesimal photon mass.  Tensor and scalar 5-point functions are
directly expressed in terms of 4-point integrals \cite{Denner:2002ii}.
Tensor 4-point and 3-point integrals are reduced to scalar integrals
with the Passarino--Veltman algorithm \cite{Passarino:1979jh} as long
as no small Gram determinant appears in the reduction. If small Gram
determinants occur, two alternative schemes are applied
\cite{Denner:2005nn}.  One method makes use of expansions of the
tensor coefficients about the limit of vanishing Gram determinants and
possibly other kinematical determinants.  In the second, alternative
method we evaluate a specific tensor coefficient, the integrand of
which is logarithmic in Feynman parametrization, by numerical
integration. Then the remaining coefficients as well as the standard
scalar integral are algebraically derived from this coefficient.  The
results of the two different codes, based on the different methods
described above are in good numerical agreement.

Since corrections due to the self-interaction of the Higgs boson
become important for large Higgs-boson masses, we have included the
dominant two-loop corrections to the decay ${\rm H}\to VV$
proportional to $G_\mu^2 M_{\rm H}^4$ in the large-Higgs-mass limit
which were calculated in Ref.~\cite{Ghinculov:1995bz}.

The matrix elements for the real photonic corrections are evaluated
using the Weyl--van der Waerden spinor technique as formulated in
Ref.~\cite{Dittmaier:1998nn} and have been checked against results
obtained with {\sl Madgraph} \cite{Stelzer:1994ta}.  The soft and
collinear singularities are treated both in the dipole subtraction
method following
Ref.~\cite{Dittmaier:2000mb} and in the phase-space slicing method
following Ref.~\cite{Bohm:1993qx}.  For the calculation of
non-collinear-safe observables we use the extension of the subtraction
method introduced in Ref.~\cite{Bredenstein:2005zk}.  Final-state
radiation beyond ${\cal O}(\alpha)$ is included at the
leading-logarithmic level using the structure functions given in
Ref.~\cite{Beenakker:1996kt} (see also references therein).

The phase-space integration is performed with Monte Carlo techniques.
One code employs a multi-channel Monte Carlo generator similar to the
one implemented in {\sl RacoonWW} \cite{Denner:1999gp} and {\sl
  Coffer}$\ga\ga$ \cite{Bredenstein:2005zk,Bredenstein:2004ef}, the
second one uses the adaptive integration program {\sl VEGAS}
\cite{Lepage:1977sw}.

\section{NUMERICAL RESULTS}

We use the $G_\mu$ scheme, i.e.\ we define the electromagnetic
coupling by $\alpha_{G_\mu}={\sqrt{2}G_\mu M_{\rm W}^2 s_{\rm
    w}^2}/{\pi}$.  Our lowest-order results include the ${\cal
  O}(\alpha)${}-corrected width of the gauge bosons.  In the
distributions the photon has been recombined with the nearest charged
fermion if the invariant mass of the photon--fermion pair is below
5\,{\rm GeV}.  More details about the setup, all input parameters, and
more detailed results are provided in Ref.~\cite{Bredenstein:2006rh}.

In Table~\ref{tab:width} the partial decay width including ${\cal
  O}(\alpha)$ corrections is shown for different decay channels and
different values of the Higgs-boson mass.
\begin{table*}
\caption{Partial decay widths for $\PH\to4\,$leptons including 
${\cal O}(\alpha)$ and ${\cal O}(G_\mu^2\MH^4)$ corrections and corresponding
relative corrections for various  decay channels and different Higgs-boson
masses (taken from Ref.~\cite{Bredenstein:2006rh}).} 
\centerline{
\begin{tabular}{|c|c|c|c|c|c|c|c|}
\hline
& $\MH[\GeV]$ & \multicolumn{2}{c|}{$140$} 
 & \multicolumn{2}{c|}{$170$} 
 & \multicolumn{2}{c|}{$200$} 
 \\
\hline
\hline
$\PH\;\to$& & $\Gamma [\MeV]$ & $\delta[\%]$ & 
$\Gamma [\MeV]$ & $\delta[\%]$ & $\Gamma [\MeV]$ & $\delta[\%]$\\
\hline
$\mathrm{e^- e^+}\mu^-\mu^+$                                     
 & corrected
 &    0.0012628(5)
 & 2.3
 &    0.020162(7)
 & 2.7
 &    0.8202(2)
 & 4.4
  \\
 &
 lowest order
 &    0.0012349(4)
 &
 &    0.019624(5)
 &
 &    0.78547(8)
 &
  \\\hline
${l^- l^+}{l^- l^+}$                               
 & corrected
 &    0.0006692(2)
 & 2.1
 &    0.010346(3)
 & 2.7
 &    0.41019(8)
 & 4.4
  \\
 $l=\Pe,\mu$ &
 lowest order
 &    0.0006555(2)
 &
 &    0.010074(2)
 &
 &    0.39286(4)
 &
  \\\hline
$\nu_\mathrm{e}\mathrm{e^+}\mu^-\bar\nu_{\mu}$                   
 & corrected
 &    0.04807(2)
 & 3.7
 &    4.3109(9)
 & 6.2
 &   12.499(3)
 & 5.0
  \\
 &
 lowest order
 &    0.04638(1)
 &
 &    4.0610(7)
 &
 &   11.907(2)
 &
  \\\hline
$\nu_l {l^+} {l^-}\bar\nu_{l}$     
 & corrected
 &    0.04914(2)
 & 3.7
 &    4.344(1)
 & 6.1
 &   14.133(3)
 & 5.0
  \\
 $l=\Pe,\mu$ &
 lowest order
 &    0.04738(2)
 &
 &    4.0926(8)
 &
 &   13.458(2)
 &
  \\\hline
\end{tabular} }
\label{tab:width}
\end{table*}
In parentheses the statistical error of the phase-space integration is
indicated, and $\delta=\Ga/\Ga_0-1$ labels the relative corrections.
The first two channels, $\mathrm{e^- e^+}\mu^-\mu^+$ and ${l^-
  l^+}{l^- l^+}$, $l=\Pe,\mu$, result from the decay
$\PH\to\PZ\PZ\to4f$. The partial widths for only electrons or muons in
the final state are equal in the limit of vanishing external fermion
masses, since for collinear-safe observables, such as the partial
widths, the fermion-mass logarithms cancel.  The width for $\PH\to{l^-
  l^+}{l^- l^+}$ is smaller by about
a factor 2, because it gets a factor $1/4$ for identical particles in
the final state and it proceeds in lowest order via two Feynman
diagrams that are related by the exchange of two outgoing electrons
and that have only a small interference.  The channel
$\nu_\mathrm{e}\mathrm{e^+}\mu^-\bar\nu_{\mu}$ results from the decay
$\PH\to\PW\PW\to4f$, while the last channel $\nu_l l^+ l^-\bar\nu_l$
receives contributions from both the decay into W and into Z~bosons.

As can be seen from the table, the relative corrections to
$\mathrm{e^- e^+}\mu^-\mu^+$ and ${l^- l^+}{l^- l^+}$ practically
coincide, demonstrating that the corrections are not significantly
influenced by the interference between the lowest-order diagrams that
exist for ${l^- l^+}{l^- l^+}$.  Similarly, there is hardly any
difference between the corrections to
$\nu_\mathrm{e}\mathrm{e^+}\mu^-\bar\nu_{\mu}$ and $\nu_l l^+
l^-\bar\nu_l$.  Plots showing the dependence of the partial decay
widths on $\MH$ in the range $\MH=120{-}700\GeV$, together with the
corresponding corrections, can be found in
Ref.~\cite{Bredenstein:2006rh}.

As examples for differential cross-sections we consider distributions
in angles between outgoing charged leptons. We note that the applied photon 
recombination does not significantly change the angular distributions,
because recombining soft or collinear photons with fermions
does not change their directions significantly.

\begin{figure*}
\setlength{\unitlength}{1cm}
\centerline{
\begin{picture}(7.7,8)
\put(-1.7,-14.5){\includegraphics{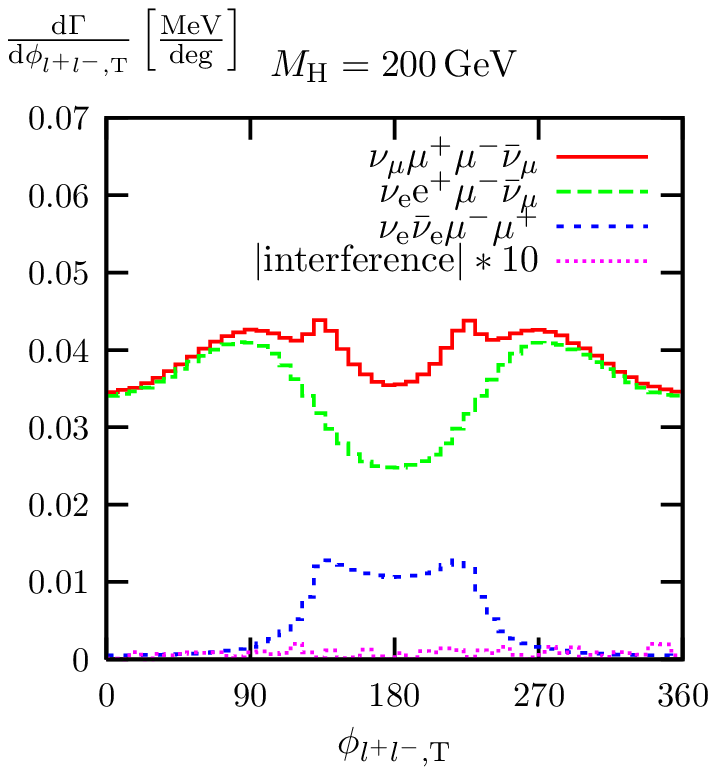}}
\end{picture}
\begin{picture}(7.5,8)
\put(-1.7,-14.5){\includegraphics{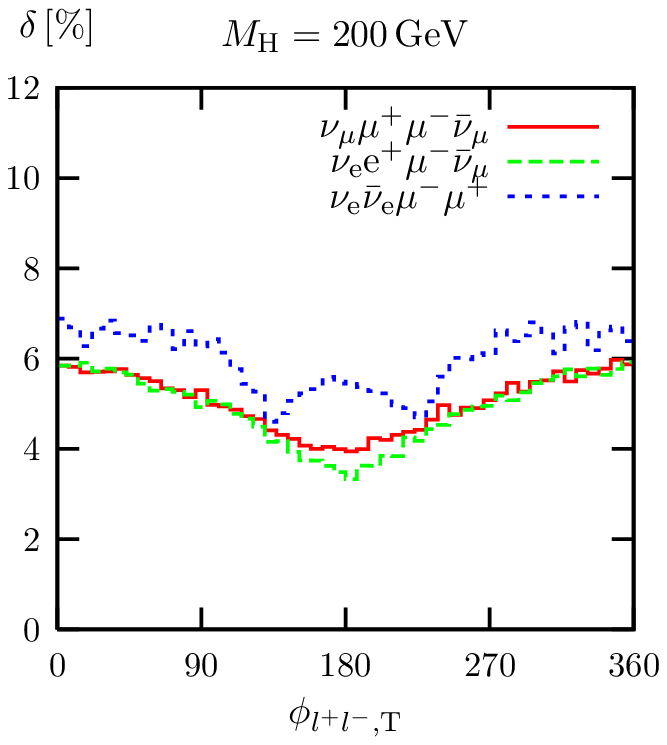}}
\end{picture} }
\vspace*{-3em}
\caption{Corrected distribution in the transverse angle between the 
  charged leptons (l.h.s.)\ and 
  corrections (r.h.s.)\ 
  in the decays of the type $\PH\to\nu_l l^+l^{\prime-}\bar\nu_l$
  for $\MH=200\GeV$.
  We define ``$|$interference$|$ ${}=
  |(\nu_\mu\mu^+\mu^-\bar\nu_\mu)
  -(\Pne\Pep\Pmum\bar\nu_\mu)
  -(\nu_\Pe\bar\nu_\Pe\mu^-\mu^+)|$''.
}
\label{fig:phitr}
\end{figure*}

In the decay $\PH\to\nu_l l^+l^{\prime-}\bar\nu_{l'}$ neither the
Higgs-boson nor the W-boson momenta can be reconstructed from the
decay products.  
Thus, angular distributions can
only be studied upon including the Higgs-production process.  
If the Higgs boson was, however, produced
without transverse momentum, or if the transverse momentum was known,
the angle between $l^+$ and $l^{\prime-}$ in the plane perpendicular
to the beam axis could be studied without knowledge of the production
process.  We define the transverse angle between $l^+$ and
$l^{\prime-}$ in the frame where ${\bf k}_{\PH,\mathrm{T}}=0$ as
\beqar
\cos\phi_{l^+l^-,\mathrm{T}} &=& 
\frac{{\bf k}_{l,\mathrm{T}}\cdot{\bf k}_{l',\mathrm{T}}}
{|{\bf k}_{l,\mathrm{T}}|{|\bf k}_{l',\mathrm{T}}|},
\nn \\
\mathrm{sgn}(\sin{\phi_{l^+l^-,\mathrm{T}} }) 
       &=& \mathrm{sgn}\{{\bf e}_z\cdot({\bf k}_{l,\mathrm{T}}\times{\bf k}_{l',\mathrm{T}})\},
\nonumber
\eeqar
where ${\bf k}_{l^{(\prime)},\mathrm{T}}$ are the 
fermion momenta transverse w.r.t.\ the unit vector ${\bf e}_z$, which could be
identified with the beam direction of a Higgs production process.
The corresponding distribution, together with the influence of the 
corrections, is shown in Figure~\ref{fig:phitr}. 
For $\MH>2\MZ$
the interference between the WW- and ZZ-initiated diagrams turns out
to be negligible, and the corrections look similar for the different
final states, in particular for those that result from
virtual W~bosons.

\begin{figure*}
\setlength{\unitlength}{1cm}
\centerline{
\begin{picture}(7.7,8)
\put(-1.7,-14.5){\includegraphics{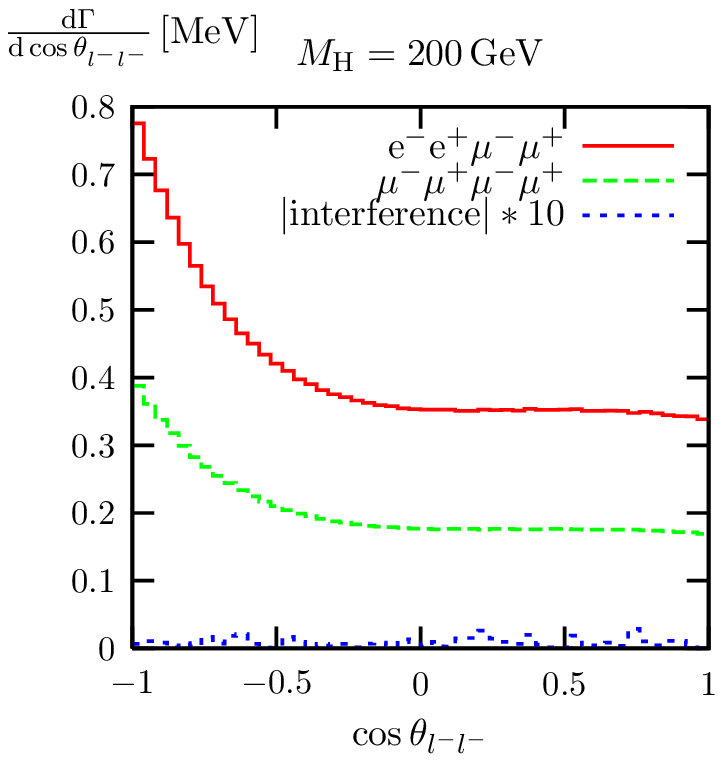}}
\end{picture}
\begin{picture}(7.5,8)
\put(-1.7,-14.5){\includegraphics{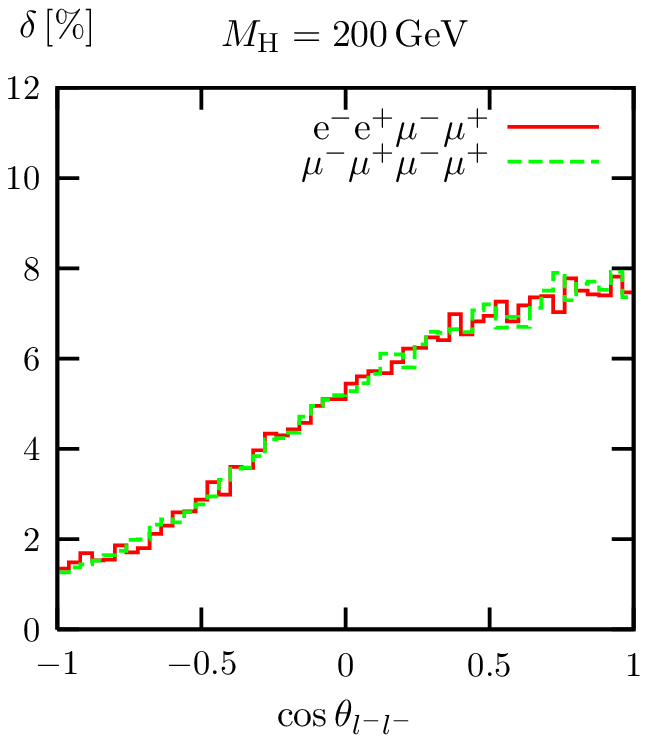}}
\end{picture} }
\vspace*{-3em}
\caption{Corrected distribution in the angle between $l^-$ and $l^{\prime-}$
  in the decays $\PH\to l^-l^+l^{\prime-}l^{\prime+}$
  (l.h.s.)\ and 
  corrections (r.h.s.)\ for $\MH=200\GeV$.
  We define ``$|$interference$|$ ${}=
  |2(\mu^-\mu^+\mu^-\mu^+)-(\Pem\Pep\mu^-\mu^+)|$''.
  }
\label{fig:th13}
\end{figure*}

Finally, we investigate the distribution of the angle between 
like-sign leptons in the decays $\PH\to\Pem\Pep\mu^-\mu^+$ and
$\PH\to\mu^-\mu^+\mu^-\mu^+$ in the rest frame of the Higgs boson.
Figure~\ref{fig:th13} shows that the
corrections tend to reduce the well-known enhancement in forward
direction.
Interference terms in the two Born diagrams for
$\PH\to\mu^-\mu^+\mu^-\mu^+$ are very small
for $\MH>2\MZ$, and the
relative corrections for the final states $\mu^-\mu^+\mu^-\mu^+$ and
$\Pem\Pep\mu^-\mu^+$ practically coincide.

\section{CONCLUSIONS}

We have presented further results of the generator {\sl PROPHECY4f\/}
which calculates the complete
electroweak ${\cal O}(\alpha)$ radiative corrections to the 
Higgs-boson decays ${\rm H}\to\PZ\PZ/\PW\PW\to 4f$.
Particular attention has been paid to 
different final states with identical 
charges, such as
$\Pep\Pem\mu^+\mu^-$ and $\mu^+\mu^-\mu^+\mu^-$.


\end{document}